\documentstyle[aps,preprint]{revtex}

\begin{document}

\bibliographystyle{prsty}

\draft

%\wideabs{

\title{  Hall anomaly in mixed state of superconductors and vortex dynamics  }
\author{ P. Ao }
\address{ Departments of Mechanical Engineering and Physics, 
          University of Washington, Seattle, WA 98195, USA }
%\address{ }
\date{ March 31, 2004 }
%\date{\today}

\maketitle

%\begin{abstract}
%\end{abstract}

%\pacs{PACS numbers: }

%}

%\maketitle

With Abrikosov shared 2003 Nobel prize in physics, everybody in the field should breathe a sign of satisfaction. Perhaps time is suitable to say a few words on vortex dynamics, yes, dynamics, and on the Hall anomaly.   

Let us admit that based on vortex many-body effect there has been a tremendous amount of theoretical progress in explaining the longitudinal resistivity and related phenomena in the mixed state. The work of the group with Larkin as the spiritual leader is the monument \cite{manybody}. Let us also admit that there has been a miserable failure in explaining the transverse resivitity, often called the anomalous Hall effect, or Hall anomaly in short. At the heart of this failure are the numerous attempts to explain the Hall anomaly based on various independent vortex dynamics theories, I believe. Because of this failure, there have been many mistakes made by numerous prominent researchers on vortex dynamics, for example, Kopnin, Sonin, Volovik, ...  .  Some of their mistakes have been corrected in a recent publication \cite{az}.

Though the present author has long been advocated the view that both longitudinal and transverse resistivities in the mixed state are dominated by vortex many-body effect.
Indeed, many concrete predictions have been made on the Hall anomaly based on such consideration \cite{ao95}. This vortex many-body view has been, however, very slow to get accepted, though its physics picture appears natural and straightforward. For example, with pinned Abrikosov lattice the excitations, such as vacancies and interstitials as well as more complicated ones, can easily lead to the Hall anomaly and various scaling laws \cite{ao95}. A more interesting competition between vortex interaction and pinning potential can even lead to many sign changes in the Hall effect \cite{ao96}.
All those predictions are perhaps not so `anomalous' after all if one takes an even causal look of the marvelous Hall effect in semiconductors. There, nobody would argue that the Lorentz equation for electrons should be changed: The Hall effect is an `obvious' competition between electron many-body effect and lattice (pinning).

It is very interesting for the present author to observe that this many-body point of view of Hall anomaly has been creeping into his most fierce opponents' work \cite{kv}, even though his own was not cited. For example, the Eq. (1) in Ref.\cite{kv} is an explicit statement on vortex many-body effect and additional ones are included through averaging ({\it c.f.} for example, Eq.(1) of  \cite{ao98} ).  
He remembers one of his most frustrating experiences in a beautiful Swedish afternoon. After explaining for numerous times on the vortex many-body effect for the Hall anomaly, Kopnin walked away saying that the vortex many-body effect cannot be true for Hall anomaly. Now we see it being employed in Kopnin's own work.

There are many lessons one can learn from here. It would be better to leave them to readers to draw their own conclusions.

\end{document}